# Single-scan, dual-functional interferometer for fast spatiotemporal characterization of few-cycle pulses

Y. F. Chen,[1,2,†] Z. Y. Huang,[1,2,†] D. Wang,[1,4] Y. Zhao,[1,2] J. H. Fu,[1,2] M. Pang,[1,3] Y. X. Leng,[1,3,5] and Z. Z. Xu[1,6]

[1]*State Key Laboratory of High Field Laser Physics and CAS Center for Excellence in Ultra-intense Laser Science, Shanghai Institute of Optics and Fine Mechanics (SIOM), Chinese Academy of Sciences (CAS), Shanghai 201800, China*
[2]*Center of Materials Science and Optoelectronics Engineering, University of Chinese Academy of Sciences, Beijing 100049, China*
[3]*Hangzhou Institute for Advanced Study, Chinese Academy of Sciences, Hangzhou 310024, China*
[4]*e-mail: wangding@siom.ac.cn*
[5]*e-mail: lengyuxin@mail.siom.ac.cn*
[6]*e-mail: zzxu@mail.shcnc.ac.cn*

**Accurate and fast characterization of spatiotemporal information of high-intensity, ultrashort pulses is crucial in the field of strong-field laser science and technology. While conventional self-referenced interferometers were widely used to retrieve the spatial profile of the relative spectral phase of pulses, additional measurements of temporal and spectral information at a particular position of the laser beam were, however, necessary to remove the indeterminacy, which increases the system complexity. Here we report an advanced, dual-functional interferometer that is able to reconstruct the complete spatiotemporal information of ultrashort pulses with a single scan of the interferometer arm. The set-up integrates an interferometric frequency-resolved optical gating (FROG) with a radial shearing Michelson interferometer. Trough scanning one arm of the interferometer, both cross-correlated FROG trace at the central part of the laser beam and delay-dependent interferograms of the entire laser profile are simultaneously obtained, allowing a fast 3-dimensional reconstruction of few-cycle laser pulses.**

Ultrashort laser pulses with temporal durations down to the few-cycle regime have now been widely used in a variety of research areas, including ultrafast spectroscopy [1], high harmonic generation [2] and laser plasma accelerators [3]. Several techniques, such as frequency-resolved optical gating (FROG) [4], spectral phase interferometry for direct electric-field reconstruction (SPIDER) [5] and D-scan [6], have been previously developed for characterizing the temporal performance of few-cycle laser pulses. None of them, however, can be used to resolve spatial information of the laser profile. Accurate and fast measurement of the spatiotemporal information of the laser pulses is essential for the development of high-intensity lasers and the applications of those lasers [7]. On one hand, the correlations between the spatial and temporal characteristics of ultrashort laser pulses, called spatiotemporal couplings (STCs), are ubiquitous, generally leading to broadening of both the focal spot size and pulse duration, and therefore leading to drop-off of the laser peak intensity at focus. On the other hand, as a prerequisite accurate measurement of spatiotemporal information of pulses renders the capability of manipulating structured laser beam, resulting in flying focus spots [8] and advanced control of light-matter interactions [7].

The spatiotemporal measurement of ultrashort pulses requires a 3-dimensional (3D) diagnosis of the light field $E(x,y,t)$. The spatial encoded arrangement (SEA) technique, like SEA-SPIDER [9], uses an imaging spectrometer to perform the spatial measurement, however it can only resolve the transverse information of laser pulses along the input slit. Fiber-assisted spatial resolved diagnosis, like STARFISH technique [10], requires a reference beam and 2-dimensional scan of the fiber end. Moreover, in some recently-developed schemes the Fourier transformation of pulse field into the spatio-spectral domain was used to describe the 3D light field, expressed as $\tilde{E}(x,y,\omega) = \sqrt{S(x,y,\omega)} \exp[i\varphi(x,y,\omega)]$, where $S(x,y,\omega)$ is the spatio-spectral intensity and $\varphi(x,y,\omega)$ is the wavefront phase. For example, in HAMSTER technique [11] the spatio-spectral pulse wavefront was measured using a Hartmann wavefront sensor with a programmable spectral filter. More recently, the total E-field reconstruction using a Michelson interferometer temporal scan (TERMITES) technique [7] was used to perform the first complete spatiotemporal reconstruction of the 100 TW laser pulses. However, this reconstruction relied on an additional measurement using Wizzler technique [12], which has to be applied to obtain temporal information of the pulses at the central part of the laser beam as the basis, increasing the complexity of the measurement procedure.

In this letter, we demonstrate a modified self-referenced spatio–temporal characterization set-up termed TERMITES-X-FROG, which can be used to measure simultaneously both the spatio-spectral wavefront of the entire laser profile and the temporal information at the central part of the laser beam. Using this set-up, accurate and fast 3D reconstruction of few-cycle laser pulses can be performed over a single scan of the interferometer arm.

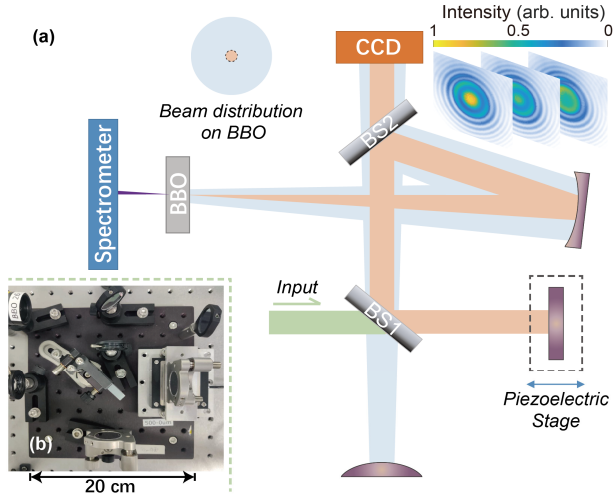

Fig. 1. (a) Schematic diagram of the measurement device. The reflecting surface of the beam splitter (BS) is marked as black. (b) Photo of the device.

Figure 1(a) shows the schematic diagram of the measurement device including the wavefront measurement and temporal diagnosis. All optical components are mounted on a 20×20 cm breadboard, as shown in Fig. 1(b). From the perspective of wavefront measurement, the principle of the device is similar as TERMITES technique. They both use an improved Michelson interferometer as a spatially resolved Fourier transform spectroscopy for multispectral wavefront reconstruction. In the experiments, the input pulses were first split into two replicas on the front surface of a coated beam splitter 1 (BS1) with a thickness of 6 mm. One replica (unknown beam, marked as blue) is radially sheared by a convex mirror, and the other replica (reference beam, marked as orange) is reflected by the mirror mounted on a piezoelectric stage (P-625.1CD from PI). Unlike TERMITES, the convex mirror was mounted on the breadboard instead of the moving stage, thus providing a constant spot magnification during the measurement process. It should be noted that the radial shearing beam passes through BS1 only once while the reference beam passes through BS1 three times, which results in more dispersion in the reference beam compared to the radial shearing beam. Then, the two beams were recombined, and the combined beams were further split on another beam splitter (BS2). When the two beams overlap in time, the transmitted beams will produce a radial shearing interference pattern. Through scanning the relative delay between the two beams, we can obtain a set of interference patterns (see the upper right corner of Fig. 1), which were captured by a charge-coupled device (CCD).

The interference patterns with respect to the delay can be used to reconstruct the wavefront in the pulses. However, measuring the wavefront at different frequencies results in the phase $\varphi(x, y, \omega)$ up to an arbitrary spectral phase function. Therefore, an additional temporal measurement is required to disambiguate the arbitrary frequency-dependent phase function. Interestingly, in our device, we innovatively integrate temporal diagnosis into the wavefront measurement without an additional temporal measurement. In Fig. 1(a), the beams reflected from BS2 were focused on a 10-μm-thickness beta-barium borate (BBO) crystal for sum-frequency generation (SFG). We used a fiber-based spectrometer (HR2000+ from Ocean Optics) to collect the delay-dependent SFG spectrogram during scanning.

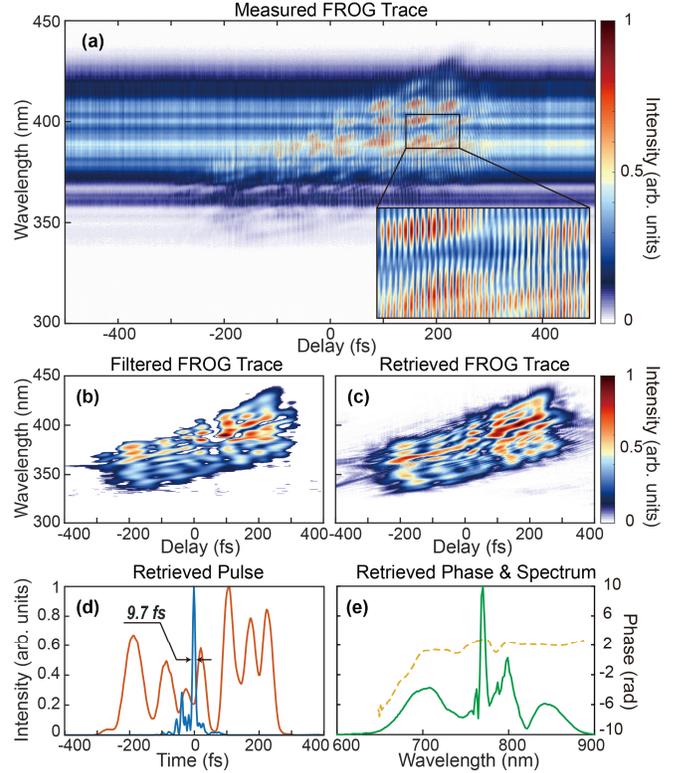

Fig. 2. Sub-10-fs ultrashort laser pulse retrieval using X-FROG. (a) Interferometric FROG trace measured by fiber spectrometer after BBO. The insert of panel (a) magnifies the interferometric modulations. (b) FROG trace with interference terms and background noise filtered out. (c) FROG trace retrieved by ePIE-FROG algorithm. (d) Retrieved temporal profiles of the two beams, the 9.7-fs beam is marked as light blue in Fig. 1(a), the other beam is marked as orange. (e) Retrieved spectral intensity and phase, corresponding to the 9.7-fs beam.

The recorded SFG spectrogram is similar to the X-FROG trace, but with interferometric modulations [see Fig. 2 (a)], which can be expressed as:

$$I(\omega, \tau) = \left| \int \left[ E_{centroid}(t) + E_R(t - \tau) \right]^2 e^{-i\omega t} dt \right|^2, \quad (1)$$

where $E_{centroid}(t)$ is the temporal pulse field at the centroid of radial shearing beam, $E_R(t - \tau)$ is the temporal field of reference beam at the focus, and $\tau$ is the relative delay between the two beams. The input pulses in the measurement device were generated from the ultrashort pulse-compression stage. In this stage, we employed the 1-m-long hollow-core fiber (HCF) with a core diameter of 250 μm and filled with ~0.58 bar Ar gas. The ~0.3 mJ, ~45 fs initial pulses from the Ti:sapphire laser system were coupled into the HCF to achieve the spectral broadening, and several pairs of chirped mirrors were used to compensated the pulses down to few-cycle range. The details about the pulse-compression stage are shown in [13-15].

Through Fourier-transforming the spectrogram and filtering the baseband [16] with a super-Gaussian window, we can obtain the X-FROG trace related to the two beams, as shown in Fig. 2(b). The background noise was eliminated by subtracting the signal at the edge where the two beams no longer overlap. Here we used the extended ptychographic iterative engine (ePIE) FROG algorithm [17] to retrieve the two pulses. In particular, the algorithm can be extended to fully characterize two different ultrashort laser pulses simultaneously, which is called the blind FROG algorithm. The retrieved FROG trace after 300 iterations is shown in Fig. 2(c). Figure 2(d) shows the retrieved temporal profiles of $E_{centroid}(t)$ (blue solid line) and $E_R(t)$ (orange solid line), respectively. We can observe that these two pulses exhibit different temporal distributions, which is due to their different dispersions obtained from BS1. The pulse duration at the centroid of radial shearing beam was measured to be 9.7 fs. In the experiment, the pulses before launching into the measurement device was pre-compensated the single-pass dispersion induced by BS1. Thus the measured pulse duration is the true width of the input pulses. In Fig. 2(e), the retrieved spectral intensity and phase of $E_{centroid}(t)$ are plotted as green solid line and yellow dashed line, respectively.

The next step is to extract wavefront through the radial shearing interference pattern captured on CCD. It should be note that the interferometric portion of the signal at any given point corresponds to the cross-correlation function of the two beams:

$$c(x,y,\tau) = \int E(x,y,\tau)E_R^*(x,y,t-\tau)dt, \quad (2)$$

where $E(x,y,\tau)$ and $E_R^*(x,y,t)$ are the complex fields of the unknown beam and reference beam, respectively. Through Fourier-transforming the cross-correlation function versus the delay $\tau$, we can obtain the cross-spectral density $\tilde{c}(x,y,\omega)$ of the two beams. The magnitude of $\tilde{c}(x,y,\omega)$ is written as:

$$|\tilde{c}| \propto \sqrt{S(x,y,\omega)} \times \sqrt{S_R(x,y,\omega)}. \quad (3)$$

Interestingly, the center of the convex mirror does not produce radial shear to the input beam, so we can easily obtain the spatio-spectral intensity $S(0,0,\omega)$ at the center of the input beam due to the $|\tilde{c}(0,0,\omega)| \propto S(0,0,\omega)$. Figure 3(a) shows the Fourier transform spectrum (yellow solid line) at the beam center and the measured spectrum (blue solid line) over the whole beam section using a fiber spectrometer with an integrating sphere. The difference between these two spectra implies spatio-spectral inhomogeneity in the pulses. Meanwhile, the phase angle of $\tilde{c}(x,y,\omega)$ is given by:

$$\phi_{\tilde{c}} = \varphi(x,y,\omega) + \varphi_c(x,y,\omega) - \varphi_R(x,y,\omega) - 2\varphi_{BS}(\omega), \quad (4)$$

where $\varphi_c(x,y,\omega)$ is the phase induced by the convex mirror, $\varphi_{BS}(\omega)$ is the spatially homogeneous phase induced by BS1, $\varphi(x,y,\omega)$ and $\varphi_R(x,y,\omega)$ are the phase of the unknown beam and reference beam, respectively. The phase $\varphi_c(x,y,\omega)$ and $\varphi_{BS}(\omega)$ can be removed numerically, so that only the phase difference between unknown beam and reference beam remains.

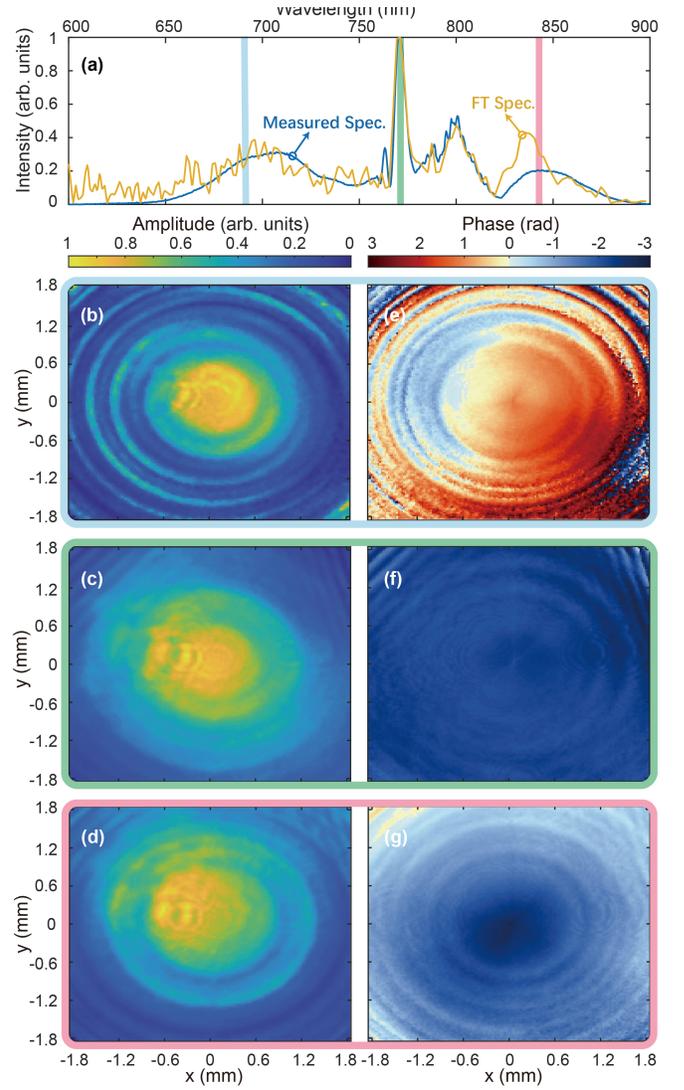

Fig. 3. Frequency-resolved spatial properties of the beam on the measurement plane. (a) The spectrum (blue solid line) measured by a spectrometer and the Fourier transform spectrum (yellow solid line) at the center of the input beam. The color bars indicate the wavelengths 691 nm, 771 nm, and 843 nm, respectively. (b)-(g) Reconstructed amplitude [(b)-(d)] and phase [(e)-(g)] profiles of the beam at three different wavelengths, corresponding to the three color bars in panel (a).

The amplitude relationship and phase relationship between the two beams are established by Eqs. (3) and (4). In consideration of the radial shearing relationship between the two beams, a simple iterative algorithm [7] was applied to retrieve the spatio-spectral amplitude $\sqrt{S(x,y,\omega)}$ and the wavefront phase $\varphi(x,y,\omega)$. Once the iterative algorithm converged, the unknown phase $\varphi(0,0,\omega)$ at the center of the beam was corrected by the phase profile [yellow dashed line in Fig. 2(e)] retrieved through the X-FROG technique. Figures 3(b)–3(d) show the reconstructed amplitude profiles for three wavelengths, indicted with different color bars in Fig. 3 (a). Obviously, amplitude distribution is highly related with wavelength, because the test pulse is the output of the

HCF, where the beam divergence depends strongly on the wavelength. Moreover, due to the cylindrical structure of the HCF, all amplitude profiles are rotationally symmetric. Chromatic phase couplings can be revealed by the reconstructed phase profiles shown in Figs. 3(e)–3(g). Although some random phase jumps can be found in the corners, the phase distributions are flat and smooth for these wavelengths. In addition, the modulation in Fig. 3 (e) has a similar spatial shape to Fig. 3 (b), which is caused by the numerical instability in the low-amplitude region.

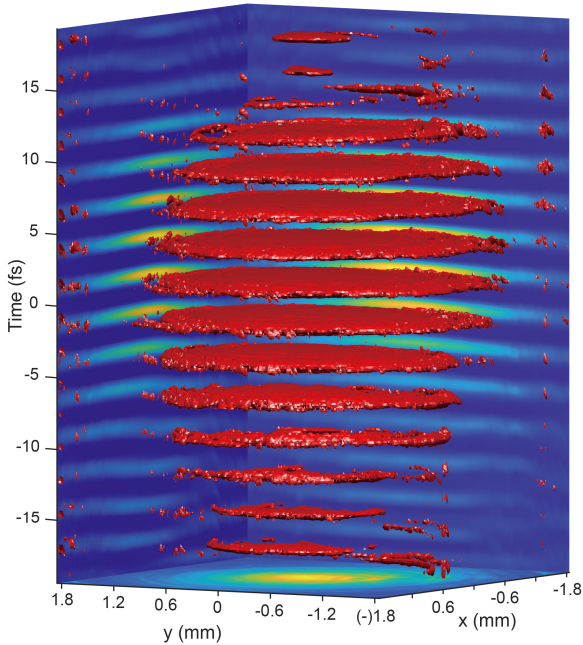

Fig. 4. 3D spatiotemporal reconstructions of the E field of the few-cycle pulses (isosurface at 20% of the peak value).

To this end, the pulse field $\tilde{E}(x,y,\omega)$ in spatio-spectral domain can be retrieved completely without ambiguity due to the accurate spectral phase at the centroid of the beam obtained through ePIE-FROG algorithm. Using the inverse Fourier transform, we can obtain the complete 3D spatiotemporal distribution of the 9.7 fs pulse field $E(x,y,t)$ [marked as red in Fig. 4] with its isosurface set to 20% of the peak field. As shown in Fig. 4, the side panels present different projections of the field $E(x,y,t)$ through integrating the pulse field along one of the three coordinates. The reconstructed 3D E field provides all information about the pulse structure, the underlying STCs, and any possible changes in pulse duration across the beam section. It is clear that the measured few-cycle pulses do not suffer from major STCs, and the duration is quite homogeneous.

In conclusion, we demonstrated that the modified self-reference spatiotemporal measurement (TERMITES-X-FROG) set-up, reported here, can be used to achieve accurate and fast spatiotemporal characterization of few-cycle pulses. In the measurements, interferograms captured by CCD camera versus the delay between the radial shearing and the reference beam were used to reconstruct the spatial wavefront of the laser beam. Meanwhile, the collinear X-FROG trace measured by a fiber-based spectrometer, retrieved by the ePIE-FROG algorithm, was used to reconstruct the spectral phase at the centroid of the laser beam. Using these results, the complete spatiotemporal information of the 9.7-fs pulses was successfully reconstructed over a single scan of the interferometer arm. The obtained 3D pulse structure show that the few-cycle pulses out of a gas-filled HCF compressor have little STCs, corresponding to a rather homogeneous pulse profile. This compact interferometer set-up, as an elegant integration of X-FROG and TERMITES techniques, provides a practical and efficient means of measuring 3D spatiotemporal information of few-cycle pulses, and may find many potential applications in fields of high-intensity lasers and ultrafast optics.

**Funding.** National Key R&D Program of China (2017YFE0123700); National Natural Science Foundation of China (61925507); Program of Shanghai Academic/Technology Research Leader (18XD1404200); Strategic Priority Research Program of the Chinese Academy of Sciences (XDB1603); Major Project Science and Technology Commission of Shanghai Municipality (2017SHZDZX02).

†These authors contributed equally to this work.